\documentclass{article}
\usepackage[a4paper, left=1in, right=1in, top=1in, bottom=1in, includehead, includefoot, foot = 0pt, head=0pt]{geometry} 
\usepackage[group-separator={,},group-digits=integer]{siunitx} 
\usepackage{booktabs} 
\usepackage{enumitem} 

\usepackage[round,colon,authoryear]{natbib} 
\usepackage[pdftex,colorlinks,citecolor=blue,urlcolor=blue,linkcolor=red]{hyperref} 
\usepackage{graphicx} 
\parskip=\medskipamount 
\usepackage[small,bf,hang]{caption}	
\usepackage{amssymb} 
\usepackage{amsmath} 
\usepackage{bbm} 
\usepackage{amsbsy} 
\usepackage{mathrsfs} 
\usepackage[linesnumbered,ruled,vlined]{algorithm2e} 
\SetAlCapFnt{\small} 
\usepackage{multirow} 
\usepackage[dvipsnames]{xcolor} 
\usepackage{subcaption} 
\usepackage[hang,flushmargin]{footmisc} 
\usepackage{tikz} 
\usetikzlibrary{matrix} 
\usepackage{rotfloat}
\usepackage{comment}
\usepackage{multirow}
\usepackage{setspace}
\usepackage{eurosym}
\usepackage[normalem]{ulem}

\usepackage{authblk}

\usepackage{fancyhdr}
\begin{document}
\title{\LARGE  Credit Scoring for Good: Enhancing Financial Inclusion with Smartphone-Based Microlending}
\author[1]{Mar\'{i}a \'{O}skarsd\'{o}ttir}
\author[2] {Crisit\'{a}n Bravo}
\author[3] {Carlos Sarraute}
\author[4,5]{Bart Baesens}
\author[4]{Jan Vanthienen}

\affil[1]{Department of Computer Science, Reykjavik University, Iceland,  Email: mariaoskars@ru.is}
\affil[2]{Department of Statistical and Actuarial Sciences, University of Western Ontario, Canada, Email: cbravoro@uwo.ca}
\affil[3]{Grandata Labs, Buenos Aires, Argentina, Email: charles@grandata.com}
\affil[4]{Faculty of Economics and Business, KU Leuven, Belgium, Email: \{bart.baesens, jan.vanthienen\}@kuleuven.be}
\affil[5]{Department of Decision Analytics and Risk, University of Southampton, UK}
\date{}

\maketitle

\begin{abstract}
Globally, two billion people and more than half of the poorest adults do not use formal financial services. Consequently, there is increased emphasis on developing financial technology that can facilitate access to financial products for the unbanked. In this regard, smartphone-based microlending has emerged as a potential solution to enhance financial inclusion.
We propose a methodology to improve the predictive performance of credit scoring models used by these applications. Our approach is composed of several steps, where we mostly focus on engineering appropriate features from the user data. Thereby, we construct pseudo-social networks to identify similar people and combine complex network analysis with representation learning. Subsequently we build credit scoring models using advanced machine learning techniques with the goal of obtaining the most accurate credit scores, while also taking into consideration ethical and privacy regulations to avoid unfair discrimination. A successful deployment of our proposed methodology could improve the performance of microlending smartphone applications and help enhance financial wellbeing worldwide.
\end{abstract}

\section{Introduction}
Worldwide, large parts of the population do not have access to useful and affordable financial products and services, such as transactions, savings, credits and insurance. The World Bank reports, that around two billion people do not have a basic bank account, and is currently on a mission to enhance financial inclusion and thus reduce poverty and boost prosperity \citep{worldBank}. This has led to a global recognition of the problem as the G20 has committed to advancing financial inclusion worldwide with over 55 countries joining the mission in addition to initiatives such as the Financial Inclusion Global Initiative (FIGI) being launched. While trying to advance financial inclusion, the countries also face various challenges, such as getting to hard-to-reach parts of the population, e.g., women and the rural poor, and increasing financial literacy. Other hurdles that are of regulatory nature include lack of official IDs, devising useful and relevant financial products and creating robust frameworks for financial consumer protection \citep{worldBank}. 

One strategy to achieve this goal of enhanced financial inclusion, is to leverage digital finance technology, also known as fintech. At the same time, cell phone ownership in developing countries is increasing. Reports from sub-Saharan Africa show that the adoption of mobile phones has transformed communication \citep{pew2015cell}. The situation in other developing countries is similar \citep{de2017financial,blondel2015survey}. This global adoption of cell phones, facilitates access to financial services to more people by means of digital IDs, digitation of payments and mobile based financial services, such as credits \citep{worldBank}. An emerging trend in that area, that could be a solution to the lack of bank access in these countries, is smartphone-based microlending. Credit bureaus are teaming up with start-ups to create applications that use machine learning techniques to analyze the data stored on the phone, e.g., call activity and app usage, to score potential borrowers and assess their creditworthiness \citep{dwoskin2015lending,kharif2016no}.

Our smartphones store a lot of data that can be exploited in various ways.  To attain the goal of enhanced financial inclusion of those that need it most, the microlending applications should make the most of that data in order to reach their highest potential, i.e., to generate the most accurate credit scores.  The apps need to make the best possible decision, while conforming to ethical and privacy regulations and without unfair discrimination. Feature engineering is therefore very important, because the data can be preprocessed and transformed in various ways.

In the last years, social networks have increasingly been used for predictive analytics in various applications \citep{dhar2014prediction,verbeke2014social, van2017gotcha}. The reason for their success is that the connections between entities in a network carry a lot of useful information.  For example in the context of customer churn prediction, closest friends can influence churn \citep{oskarsdottir2017social}. However, scoring someone on who their friends are is unfair and unethical.  A possible solution is to obtain networks in a different way, e.g., by linking together people who are similar. Features extracted from these pseudo-social networks have been shown to have high predictive performance in different domains \citep{martens2016mining,lismont2018predicting}.

In this paper, we propose a multilevel approach for credit risk modeling in the context of microlending smartphone applications. We combine complex network analysis with state-of-the-art representation learning techniques to featurize a pseudo-social network representing similarities between people.  These features can subsequently be used in a credit scoring exercise and the predictive performance of various feature types compared. A successful deployment of our proposed methodology could enhance the performance of microlending smartphone applications and potentially help increase the financial wellbeing of numerous individuals worldwide.

The rest of this paper is organized as follows. In the next section we discuss related literature on credit scoring with alternative data and social networks for predictive analytics.  Then, we briefly present results of an earlier case study we performed, as a motivation for the techniques we propose for this study.  Subsequently, we give a detailed overview of our proposed methodology. Thereafter, we address concerns regarding the ethics of this research project. To conclude, we discuss the implications of our research.

\section{Related Work}
For the purpose of this paper, we distinguish two relevant fields of research: Using alternative data for credit scoring and predictive analytics with social networks.
\subsection{Alternative Data for Credit Scoring}
Credit scoring is one of the oldest applications of analytics that traditionally exploits personal and banking information together with repayment history to assess creditworthiness and thereby avoid giving credits to people that are likely to default.  In this field of analytics, the goal is to build a model that distinguishes bad borrowers from good ones in order to give credit to people who are more reliable and likely to pay back their loans. Since the accuracy and reliability of credit scoring techniques is crucial for lenders and lendees alike, their development has been a popular research topic for decades \citep{baesens2016credit}. Recent research shows that state-of-the-art techniques, which are often heterogeneous ensembles that are complex and difficult to interpret, rarely outperform the more seasoned methods such as random forests and neural networks \citep{lessmann2015benchmarking}. In order to improve credit decision-making, the focus should therefore shift to other elements of the credit scoring task, in particular to leverage innovative data sources.

Because of the numerous challenges associated with cleaning, transforming, integrating and modeling the data, it is a non-trivial and difficult task \citep{alexander2017research}. However, existing results are promising and show the potential for alternative data sources. \citet{singh2015money} analyzed the time and location of economic transactions and showed that models which incorporate spatio-temporal traits such as exploration, engagement and elasticity have higher predictive performance. In terms of social networks for peer-to-peer lending, \citet{lin2013judging} demonstrated that the online friendships of borrowers act as signals of credit quality. Furthermore, \citet{wei2015credit} developed a series of models to study the impact of tie formation and compare the accuracy of credit scores with and without network data.  They did however not evaluate the performance on real data. \citet{de2019does} studied the importance of both social and pseudo-social networks for microlending predictions, building networks using Facebook data. Their results indicated, that features from the implicit networks had more predictive power than features from explicit friends networks. Finally, \citet{ruiz2017credit} analyzed a microlending application dataset to predict creditworthiness, using only user and basic mobile usage data together with logistic regression and support vector machines. Their results showed that these applications are indeed worthwhile, although there is still place for improvement.

\subsection{(Pseudo-)Social Networks and Predictive Analytics}
Predictive analytics is a rich field in the world of data mining. It has a wide range of applications, such as in customer relationship management, marketing and credit risk modelling. Networks are being exploited more frequently for predictive modeling, commonly in form of social network analysis.  The observations in a network, the nodes, do not behave in the same way as observations in regular datasets, because they are all connected and may therefore have an effect on each other. At least two common ways exist to make predictions using information from networks. Firstly, the information about the connected nodes and links can be used to make inferences in the network directly, for example regarding age and gender of people. This is known as network learning \citep{macskassy2007classification}. Secondly, the network can be transformed into a tabular dataset by extracting features that represent the role and position of each node.  These features are subsequently used as input in binary classifiers, such as logistic regression, to make the predictions \citep{van2017gotcha}. 

The art of featurization is complex and not always a straightforward task.  Transforming a network into features which are descriptive and represent the network correctly, is difficult and highly dependent on the application. Consequently, there have been attempts at automating this process using representation learning. \citet{grover2016node2vec} proposed the technique node2vec, which generates a set of features using a two-step random walk to build paths that are subsequently fed into a skip gram model to create features that are representative of the network.  

Similar to actual social networks, implicit or pseudo-social networks, where the entities are linked together based on common features, rather than direct friendship, have been used in predictive modeling with good results.  \citet{martens2016mining} exploited pseudo-social networks in a financial setting, by linking together people based on purchase history. Furthermore, \citet{lismont2018predicting} featurized customer-product networks in the context of predicting interpurchase times of the products, which resulted in enhanced model performance.  

\section{Previous Work}
In a previous credit scoring study of people applying for credit cards, we used a combination of customer bank data and their mobile phone data. Our results showed that individual calling behavior is significantly predictive in terms of  creditworthiness \citep{oskarsdottir2019value}. This unique combination of datasets contains a year and a half of banking history of over 2 million customers as well as five consecutive months of calling activity of 90 million unique phone numbers. The mobile phone data allowed us to construct social networks based on the exchange of phone calls between people, an approach that has been proven successful in other applications, such as for churn prediction \citep{oskarsdottir2017social}.

\begin{figure}
\centering
\includegraphics[scale=0.8]{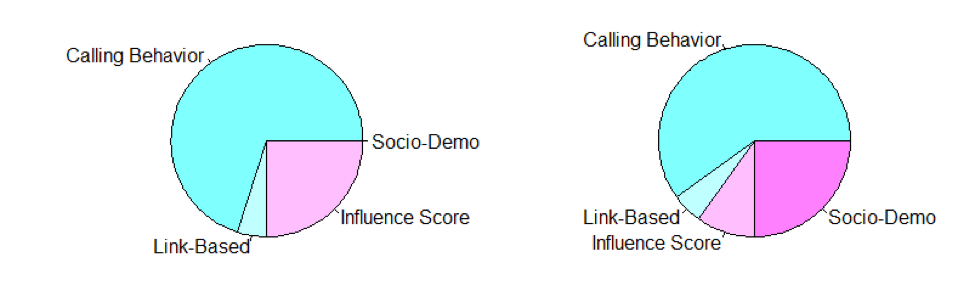}
\caption{ Statistical (left) and economic (right) feature importance\label{fig:creditpaper}}
\end{figure}

To build credit scoring models for the credit card applicants, we extracted four types of features. These were: 
\begin{enumerate}
\item
Socio-Demographic: including bank history, income and debit account behavior. 
\item 
Calling Behavior: aggregated values for number and duration of phone calls made and received on different days and at different times of the day. 
\item Link-Based: counts of the number of good and bad credit card holders in each customer's egonet. 
\item Influence Score: the scores each customer obtained after two distinct influence propagation algorithms were applied to the network. 
\end{enumerate}
 Bad credit card holders were used in lieu of information source, as explained in sub-subsection ‘Centrality Features’. Finally, we used the bank data to label the customers as defaulters and non-defaulters.

We built a credit scoring model using random forests with all the extracted features. Random forests were chosen because they are a powerful classification technique that is also capable of ranking the importance of the input features \citep{breiman2001random}. Thereby, we could identify which features were most important in terms of model performance. The models were evaluated from both a statistical and an economic perspective. In the first case, we used accuracy, i.e., the ratio of correctly classified instances, and the commonly applied area under the ROC curve (AUC) measure, which gives the probability of a randomly chosen bad customer being ranked higher by the model than a randomly chosen good customer. To measure economic performance we deployed the Expected Maximum Profit measure for credit scoring, which has the advantage of considering the expected losses and operational income generated by the loan, and is thus tailored towards the business goal of credit scoring \citep{verbraken2014development}. Moreover, when applied to credit scoring models, it facilitates computing the model’s value and allows us to identify which features are most favorable in terms of profit.

The results for the most important features can be seen in Figure \ref{fig:creditpaper}. The pie chart on the left shows the statistical feature importance and the pie chart on the right shows the economic feature importance among the 20 most important features grouped by feature type. Evidently, the calling behavior features are most important as they take up more than half of both the charts. In terms of statistical performance the socio-demographic features do not appear at all. This is an interesting result since it indicates that people's phone usage can be used as the sole data source when deciding whether they should be granted a credit.

\section{Research Questions}
In this research, we intend to address the two following research questions in the context of smartphone-based microlending:
\begin{itemize}
\item[RQ1]
 Which type of features is most predictive of creditworthiness? Various types of features can be extracted from the data and used to construct models to predict creditworthiness. Firstly, calling behavior represented by the frequency and timing of phone calls and text messages. Secondly, featurization of pseudo-social networks that are constructed by linking together similar people, e.g. based on common characteristics.  Next, influence scores resulting from default propagation techniques. Finally, many other types of features will be explored. The performance of the resulting models are compared to determine which type of features is most predictive.
\item[RQ2] 
What is the most effective way to construct and featurize the network? As described below, pseudo-social networks can be constructed in many ways and the subsequent feature engineering is also a complex task.  We will investigate which approach for these two steps is best for both the predictive performance of the credit scoring models as well as for the operational efficiency of the application. 
\end{itemize}

\section{Methodology}
Smartphones are a provenance of data that can be leveraged in various ways. In this section, we provide a detailed description of our proposed methodology for featurizing the data, constructing pseudo-social networks and building credit scoring models.
\subsection{Data Description}
For this research, we have at our disposal a dataset of a smartphone-based microlending application. Using this application, the borrowers –or users- can request a loan of a few fixed amounts. If the borrower is deemed worthy, i.e., has a good enough credit score, the loan is granted instantly. The loan term is one month, and should be repaid with interests at the end of the period in one installment.  A borrower is considered a defaulter if they do not repay the loan within two months from receiving it. We refer to borrowers who have defaulted as bad users and the borrowers who pay their loans back on time as good users. The application has access to the data that is stored on the device, such as the user’s list of apps, phone book and text messages. Due to privacy regulation, we cannot disclose more information about the data or the operating country.
\subsection{Local Features}
Much of the data on the smartphone is user specific and can be used directly for credit scoring. These local features represent information related to users as isolated entities without connections to the outside world. These are for example, socio-demographic features, such as age and residence, calling and texting behavior, and number and categories of various applications. As in the study described in section ‘Previous Work’, we make a distinction between socio-demographic features, which include basic statistics about the user and the applications on their smartphone, and behavior features, which we define as aggregated values of the number and time of made and received phone calls and text messages.
\subsection{Pseudo-Social Networks}
\begin{figure}
\centering
\includegraphics[scale=0.33]{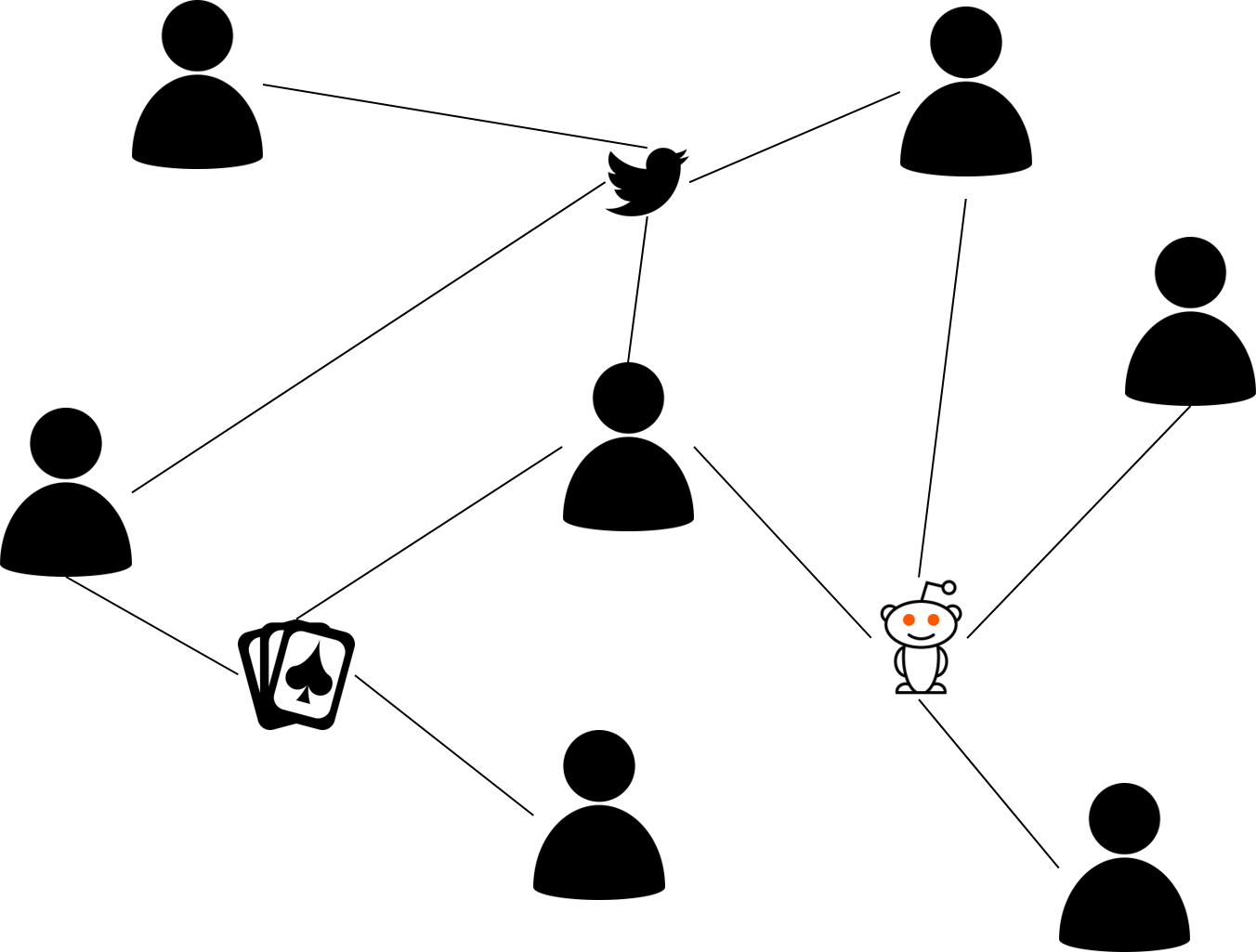}
\caption{ Bipartite Network \label{fig:bipart}}
\end{figure}

The data can be transformed to construct networks, in this case pseudo-social networks, as described below. 
In the first step of the process, we construct the pseudo-social network based on a specific common feature, e.g., being frequent users of certain apps. This is depicted in Figure \ref{fig:bipart}, which shows a network of seven users and three apps: Twitter (the bird), reddit (the alien), and a gambling app (the cards). We link a user to an app if they have it on their smartphone and use it frequently. This is a bipartite network with two types of nodes: users and apps. In this network, the links are unweighted, but we could add weights, for example to indicate the intensity of the usage. We denote this network with $N_b$.

\begin{figure}
\centering
\includegraphics[scale=0.33]{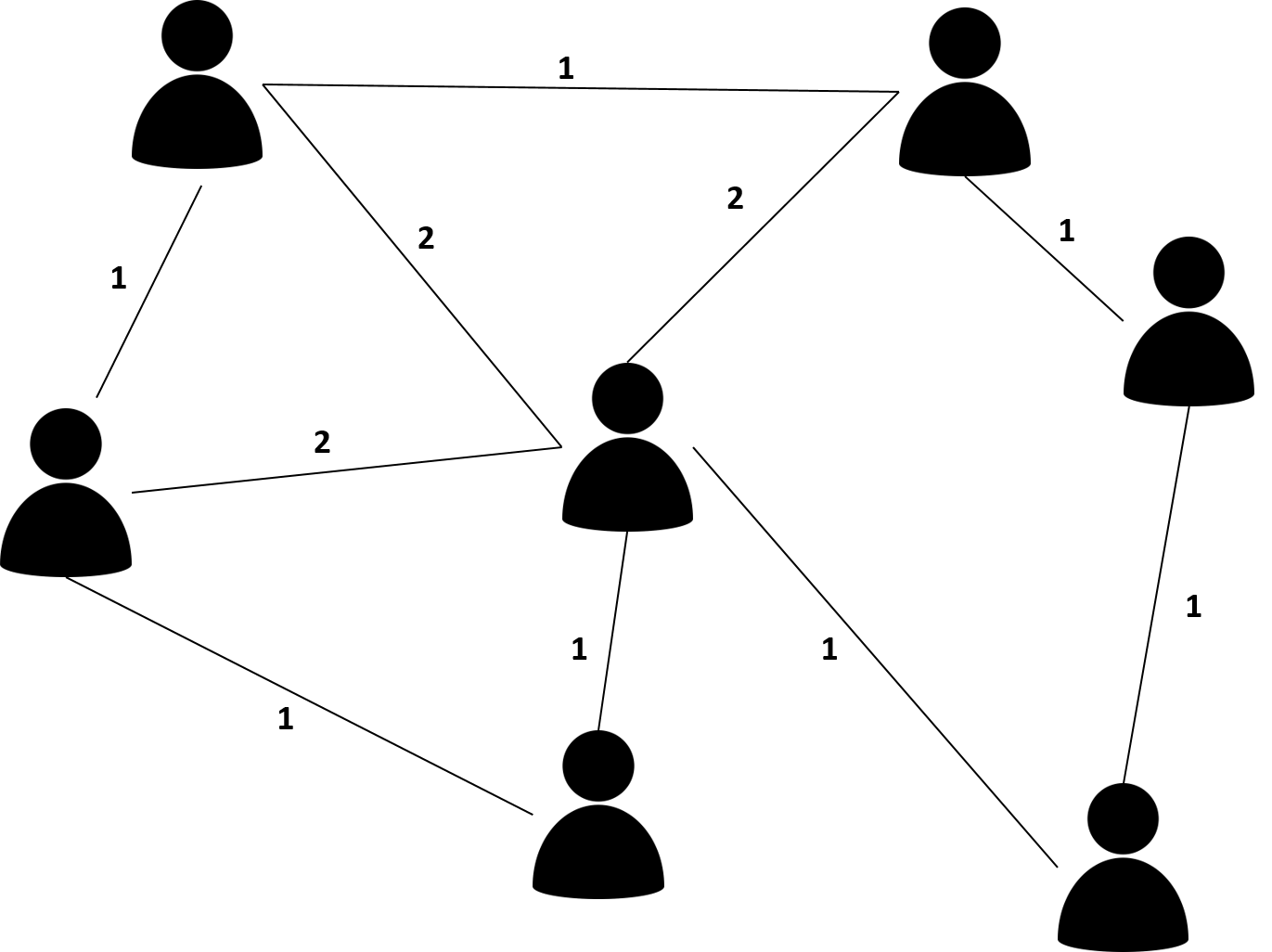}
\caption{ Unipartite Network \label{fig:unipart}}
\end{figure}

\begin{figure}
\centering
\includegraphics[scale=0.33]{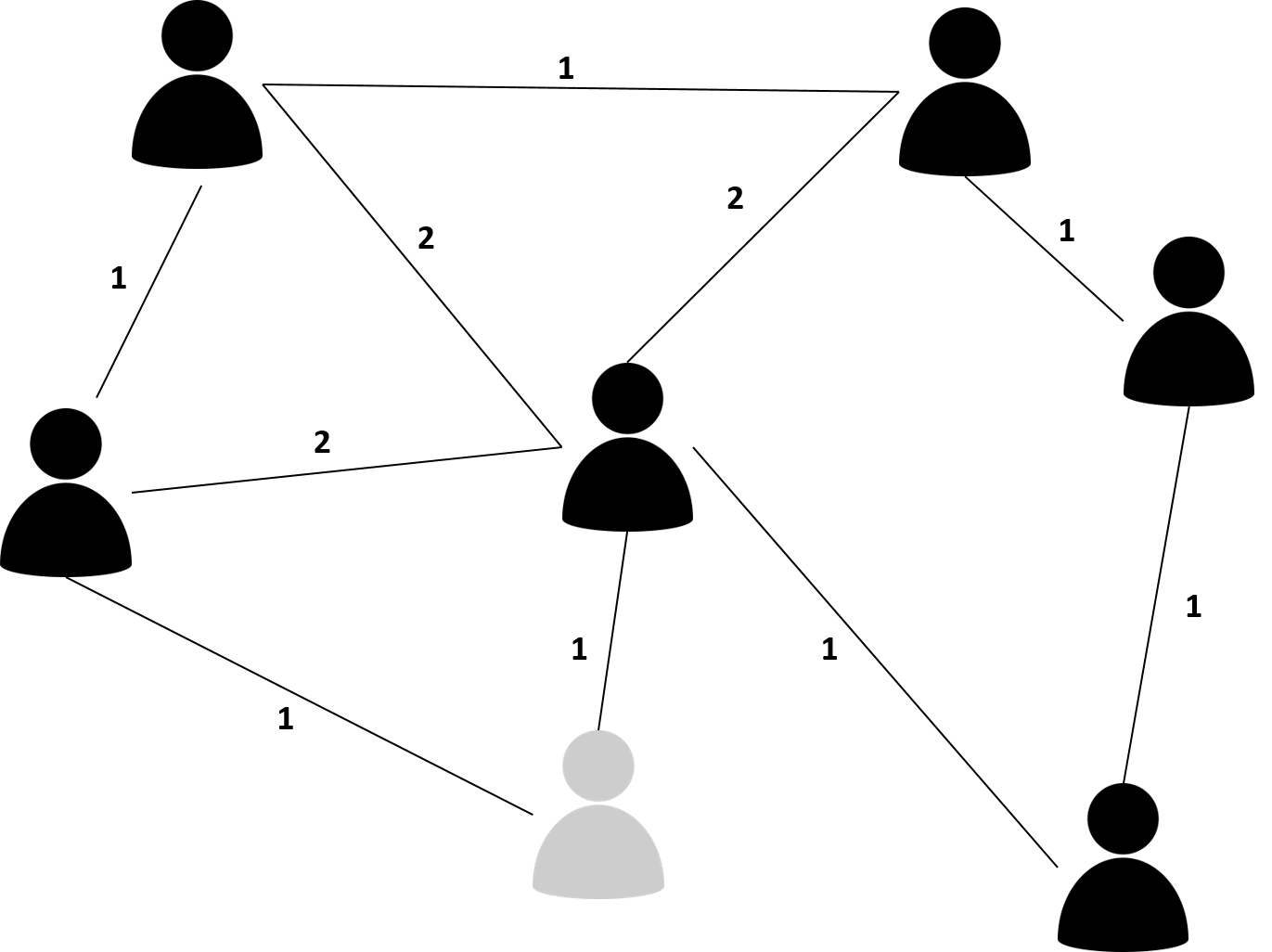}
\caption{Network with one defaulter \label{fig:unipart2}}
\end{figure}

In the second step, we convert the bipartite network to a unipartite network which we indicate with $N_u$. In this network, two users are linked if both of them had the same app on their smartphone. The weights on the links indicate the number of apps they had in common.  The network in Figure \ref{fig:unipart} shows the unipartite version of the network in Figure \ref{fig:bipart}.  

Finally, we look at a third version of the network, where we have labelled users that defaulted in the past, as seen in Figure \ref{fig:unipart2}.  In this network, the bad users are gray and the good users are black.  This allows us to extract features that take into account status of similar users, and propagate influence in the network.
\subsection{Network Featurization}
A social network can be used in predictive modeling, for example by extracting features to form a tabular dataset.  There are many types of features as described below. 
\paragraph{Neighborhood Features}
Neighborhood features are derived from a user’s neighborhood, and thus provide a characterization relative to the most similar users.  We will consider the egonet of a user, which consists of those users that are directly connected to it.  The degree of a node is the number of nodes in the egonet. If we furthermore make a distinction between good and bad users in the egonet, we get the features good degree and bad degree.  The user in the center of Figure \ref{fig:unipart2} has degree, good degree and bad degree equal to six, five and one, respectively. Other neighborhood features include number of triangles, i.e. fully connected subgraphs of three nodes, transitivity, which measures how densely connected the egonet is, and relational neighbor, the weighted average of bad nodes in the neighborhood. These features are easy to extract from the network but give a very representative image of how a user compares to the others.
\paragraph{Centrality Features}
Centrality features quantify the position and importance of a node in a network. Two common features are  closeness, i.e. the average distance to all nodes in the network, and betweenness, i.e. the number of times a node lies on the shortest path between two other nodes.

The Personalized PageRank algorithm was developed to rank the importance of webpages for search engines by simulating surfing behavior \citep{page1999pagerank}. It has also been used to measure the importance of nodes in a network \citep{van2017gotcha,lismont2018predicting}. The addition of a restart vector, or information source, facilitates higher ranking of nodes that are closer to the source. The Personalized PageRank algorithm iteratively updates the PageRank scores of the nodes in the network using the equation 
\begin{equation}\label{eq:PR}
\vec{r}=\alpha\cdot A\cdot \vec{r}+(1-\alpha)\cdot\vec{e}
\end{equation}
where $\vec{r}$ is the vector of PageRank scores, A is the adjacenty matrix of the network and $\alpha$ is a damping factor. The restart vector $\vec{e}$ indicates the information source that can be defined in various ways. The result is an influence score for each node in the network, where users who are highly exposed to the particular influence get a higher score. We compute influence scores using different definitions of the information source, e.g., the set of all bad users in the unipartite network (as \citet{van2017gotcha} did for fraud detection) or recency and frequency indicators for the apps in the bipartite network (as \citet{lismont2018predicting} did for a customer-product network). 
\paragraph{Representation Learning}
The feature extraction process can be automated using representation learning. We use a recently proposed technique called node2vec, that is able to capture the diverse connectivity patterns in networks \citep{grover2016node2vec}. It learns a mapping of the nodes to a low-dimensional space of continuous features that preserve the network topology. We use the node2vec technique to extract features for the users in the network. As the number of generated features is user-specified, we explore how many are needed to achieve accurate predictions and, in addition, we consider varying network architectures, with different definitions of weights in both the unipartite and the bipartie networks.  
\subsection{Predicting Creditworthiness}
To assess the creditworthiness of users asking for a microloan with the app, we build credit scoring models using binary classification techniques. As \citet{lessmann2015benchmarking} advised in their benchmarking study of credit scoring models, we use random forests and artificial neural networks as benchmark techniques, which is appropriate given the high dimensionality of our data. In addition, we deploy logistic regression, since it is the industry standard. Furthermore, we evaluate and compare the model performance using the commonly used AUC measure and the Brier Score as well as a recently proposed profit measure \citep{verbraken2014development}. This combination of measures allows us to assess the discriminatory ability of the credit scores, the accuracy of their probability predictions and economic impact, respectively.

To address the two research questions, we build multiple models with various combination of features to establish which type is most predictive. Furthermore we perform statistical tests to establish whether differences in performance are significant or not. In that way, we can for example compare the performance of a credit scoring model that uses calling behavior features to the performance of a model that uses features resulting from a representation learning technique. As a result, we can find which type of features or which combination of feature types is most predictive of creditworthiness.  
\section{Privacy and Ethical Concerns}
The main goal of this research is to show that smartphone-based microlending applications are a viable option for enhancing financial inclusion of the unbanked. However, personal behavior information from smartphones for profiling potential borrowers raises questions about the ethical nature of our approach. Although the data does facilitate constructing detailed and real social networks, it would be both unfair and unethical to use such explicit information when scoring the users. Instead we focus on, on the one hand, individual behavior, i.e., personal usage data stored on the device -not that of their contacts-, and, on the other hand, implicit network insights in the form of pseudo-social networks. Therefore, we leverage characteristics of similar users when featurizing the smartphone data to obtain more accurate credit scores.

Lately, there has been increased emphasis on using positive information that represents good financial behavior for credit scoring \citep{worldBank2011}.  That is exactly what microlending applications can achieve. While enhancing the financial wellbeing and quality of life, they simultaneously help people build their credit scores by means of small loans.

Because of the ethical and privacy considerations associated with our research, we have made sure that we are acting in accordance with ethics regulations. Furthermore, as the data is fully anonymized and contains no personally identifiable information, so we also comply with privacy regulations.
\section{Conclusion}
Smartphone based microlending has emerged as a feasible option to obtain credit. In this paper, we proposed a multilevel approach to investigate how to optimally leverage the data to obtain accurate credit scores. Using state-of-the-art machine learning and social network techniques, our application explores various ways of featurizing the data. Our technique is furthermore considerate of ethical and privacy concerns as it focuses on similar users and not actual social networks. By comparing the effect different features have on the credit scores, we can determine which type of features is most predictive and has the highest relevance for this specific task. The results of the previous study indicated that calling behavior alone is highly predictive of creditworthiness. In this study, we hope to find other behavioral indicators in the pseudo-social networks that can be used in credit scoring models to obtain even better model performance and thus better predictions to decide whether a loan should be granted. 

The impact of our research is manifold. First of all, our results have the potential to increase the performance of credit scoring models which will improve credit decision-making and pricing. Furthermore, the insights obtained from this research could enhance financial inclusion in developing countries by facilitating access to credit for borrowers with little or no credit history. As cell phone usage is rapidly growing, their users have the opportunity to request credits where the decision is based on data stored on the device. The benefits of these applications are substantial since they can help increase the financial wellbeing of numerous individuals worldwide.  Their success however, highly depends on using descriptive features in appropriate credit scoring models that are able to accurately assess creditworthiness. 
\bibliographystyle{apacite}
\bibliography{biblio}

\begin{thebibliography}{27}
\providecommand{\natexlab}[1]{#1}
\providecommand{\url}[1]{\texttt{#1}}
\expandafter\ifx\csname urlstyle\endcsname\relax
  \providecommand{\doi}[1]{doi: #1}\else
  \providecommand{\doi}{doi: \begingroup \urlstyle{rm}\Url}\fi

\bibitem[Alexander et~al.(2017)Alexander, Das, Ives, Jagadish, and
  Monteleoni]{alexander2017research}
Lewis Alexander, Sanjiv~R Das, Zachary Ives, HV~Jagadish, and Claire
  Monteleoni.
\newblock Research challenges in financial data modeling and analysis.
\newblock \emph{Big data}, 5\penalty0 (3):\penalty0 177--188, 2017.

\bibitem[Baesens et~al.(2016)Baesens, Roesch, and Scheule]{baesens2016credit}
Bart Baesens, Daniel Roesch, and Harald Scheule.
\newblock \emph{Credit risk analytics: Measurement techniques, applications,
  and examples in SAS}.
\newblock John Wiley \& Sons, 2016.

\bibitem[Blondel et~al.(2015)Blondel, Decuyper, and Krings]{blondel2015survey}
Vincent~D Blondel, Adeline Decuyper, and Gautier Krings.
\newblock A survey of results on mobile phone datasets analysis.
\newblock \emph{EPJ data science}, 4\penalty0 (1):\penalty0 10, 2015.

\bibitem[Breiman(2001)]{breiman2001random}
Leo Breiman.
\newblock Random forests.
\newblock \emph{Machine learning}, 45\penalty0 (1):\penalty0 5--32, 2001.

\bibitem[De~Cnudde et~al.(2019)De~Cnudde, Moeyersoms, Stankova, Tobback,
  Javaly, and Martens]{de2019does}
Sofie De~Cnudde, Julie Moeyersoms, Marija Stankova, Ellen Tobback, Vinayak
  Javaly, and David Martens.
\newblock What does your facebook profile reveal about your creditworthiness?
  using alternative data for microfinance.
\newblock \emph{Journal of the Operational Research Society}, 70\penalty0
  (3):\penalty0 353--363, 2019.

\bibitem[de~Luna~Mart{\'\i}nez(2017)]{de2017financial}
Jos{\'e} de~Luna~Mart{\'\i}nez.
\newblock Financial inclusion in malaysia--distilling lessons for other
  countries.
\newblock \emph{World Bank Working Paper}, 115155, 2017.

\bibitem[Dhar et~al.(2014)Dhar, Geva, Oestreicher-Singer, and
  Sundararajan]{dhar2014prediction}
Vasant Dhar, Tomer Geva, Gal Oestreicher-Singer, and Arun Sundararajan.
\newblock Prediction in economic networks.
\newblock \emph{Information Systems Research}, 25\penalty0 (2):\penalty0
  264--284, 2014.

\bibitem[Dwoskin(2015)]{dwoskin2015lending}
E~Dwoskin.
\newblock Lending startups look at borrowers’ phone usage to assess
  creditworthiness.
\newblock \emph{The Wall Street Journal}, 2015.

\bibitem[Grover and Leskovec(2016)]{grover2016node2vec}
Aditya Grover and Jure Leskovec.
\newblock node2vec: Scalable feature learning for networks.
\newblock In \emph{Proceedings of the 22nd ACM SIGKDD international conference
  on Knowledge discovery and data mining}, pages 855--864, 2016.

\bibitem[Kharif(2016)]{kharif2016no}
Olga Kharif.
\newblock No credit history? no problem. lenders are looking at your phone
  data.
\newblock \emph{Bloomberg}, November 2016.
\newblock URL
  \url{https://www.bloomberg.com/news/articles/2016-11-25/no-credit-history-no-problem-lenders-now-peering-at-phone-data}.

\bibitem[Lessmann et~al.(2015)Lessmann, Baesens, Seow, and
  Thomas]{lessmann2015benchmarking}
Stefan Lessmann, Bart Baesens, Hsin-Vonn Seow, and Lyn~C Thomas.
\newblock Benchmarking state-of-the-art classification algorithms for credit
  scoring: An update of research.
\newblock \emph{European Journal of Operational Research}, 247\penalty0
  (1):\penalty0 124--136, 2015.

\bibitem[Lin et~al.(2013)Lin, Prabhala, and Viswanathan]{lin2013judging}
Mingfeng Lin, Nagpurnanand~R Prabhala, and Siva Viswanathan.
\newblock Judging borrowers by the company they keep: Friendship networks and
  information asymmetry in online peer-to-peer lending.
\newblock \emph{Management Science}, 59\penalty0 (1):\penalty0 17--35, 2013.

\bibitem[Lismont et~al.(2018)Lismont, Ram, Vanthienen, Lemahieu, and
  Baesens]{lismont2018predicting}
Jasmien Lismont, Sudha Ram, Jan Vanthienen, Wilfried Lemahieu, and Bart
  Baesens.
\newblock Predicting interpurchase time in a retail environment using
  customer-product networks: An empirical study and evaluation.
\newblock \emph{Expert systems with applications}, 104:\penalty0 22--32, 2018.

\bibitem[Macskassy and Provost(2007)]{macskassy2007classification}
Sofus~A Macskassy and Foster Provost.
\newblock Classification in networked data: A toolkit and a univariate case
  study.
\newblock \emph{Journal of machine learning research}, 8\penalty0
  (May):\penalty0 935--983, 2007.

\bibitem[Martens et~al.(2016)Martens, Provost, Clark, and
  de~Fortuny]{martens2016mining}
David Martens, Foster Provost, Jessica Clark, and Enric~Junqu{\'e} de~Fortuny.
\newblock Mining massive fine-grained behavior data to improve predictive
  analytics.
\newblock \emph{MIS quarterly}, 40\penalty0 (4), 2016.

\bibitem[{\'O}skarsd{\'o}ttir et~al.(2017){\'O}skarsd{\'o}ttir, Bravo, Verbeke,
  Sarraute, Baesens, and Vanthienen]{oskarsdottir2017social}
Mar{\'\i}a {\'O}skarsd{\'o}ttir, Cristi{\'a}n Bravo, Wouter Verbeke, Carlos
  Sarraute, Bart Baesens, and Jan Vanthienen.
\newblock Social network analytics for churn prediction in telco: Model
  building, evaluation and network architecture.
\newblock \emph{Expert Systems with Applications}, 85:\penalty0 204--220, 2017.

\bibitem[{\'O}skarsd{\'o}ttir et~al.(2019){\'O}skarsd{\'o}ttir, Bravo,
  Sarraute, Vanthienen, and Baesens]{oskarsdottir2019value}
Mar{\'\i}a {\'O}skarsd{\'o}ttir, Cristi{\'a}n Bravo, Carlos Sarraute, Jan
  Vanthienen, and Bart Baesens.
\newblock The value of big data for credit scoring: Enhancing financial
  inclusion using mobile phone data and social network analytics.
\newblock \emph{Applied Soft Computing}, 74:\penalty0 26--39, 2019.

\bibitem[Page et~al.(1999)Page, Brin, Motwani, and Winograd]{page1999pagerank}
Lawrence Page, Sergey Brin, Rajeev Motwani, and Terry Winograd.
\newblock The pagerank citation ranking: Bringing order to the web.
\newblock Technical report, Stanford InfoLab, 1999.

\bibitem[{Pew Research Center}(2015)]{pew2015cell}
{Pew Research Center}.
\newblock Cell phones in africa: Communication lifeline.
\newblock \emph{Pew Research Center}, 2015.
\newblock URL
  \url{http://www.pewglobal.org/2015/04/15/cell-phones-in-africa-communication-lifeline/}.

\bibitem[Ruiz et~al.(2017)Ruiz, Gomes, Rodrigues, and Gama]{ruiz2017credit}
Saulo Ruiz, Pedro Gomes, Lu{\'\i}s Rodrigues, and Jo{\~a}o Gama.
\newblock Credit scoring in microfinance using non-traditional data.
\newblock In \emph{EPIA Conference on Artificial Intelligence}, pages 447--458.
  Springer, 2017.

\bibitem[Singh et~al.(2015)Singh, Bozkaya, and Pentland]{singh2015money}
Vivek~Kumar Singh, Burcin Bozkaya, and Alex Pentland.
\newblock Money walks: implicit mobility behavior and financial well-being.
\newblock \emph{PloS one}, 10\penalty0 (8):\penalty0 e0136628, 2015.

\bibitem[Van~Vlasselaer et~al.(2017)Van~Vlasselaer, Eliassi-Rad, Akoglu,
  Snoeck, and Baesens]{van2017gotcha}
V{\'e}ronique Van~Vlasselaer, Tina Eliassi-Rad, Leman Akoglu, Monique Snoeck,
  and Bart Baesens.
\newblock Gotcha! network-based fraud detection for social security fraud.
\newblock \emph{Management Science}, 63\penalty0 (9):\penalty0 3090--3110,
  2017.

\bibitem[Verbeke et~al.(2014)Verbeke, Martens, and Baesens]{verbeke2014social}
Wouter Verbeke, David Martens, and Bart Baesens.
\newblock Social network analysis for customer churn prediction.
\newblock \emph{Applied Soft Computing}, 14:\penalty0 431--446, 2014.

\bibitem[Verbraken et~al.(2014)Verbraken, Bravo, Weber, and
  Baesens]{verbraken2014development}
Thomas Verbraken, Cristi{\'a}n Bravo, Richard Weber, and Bart Baesens.
\newblock Development and application of consumer credit scoring models using
  profit-based classification measures.
\newblock \emph{European Journal of Operational Research}, 238\penalty0
  (2):\penalty0 505--513, 2014.

\bibitem[Wei et~al.(2015)Wei, Yildirim, Van~den Bulte, and
  Dellarocas]{wei2015credit}
Yanhao Wei, Pinar Yildirim, Christophe Van~den Bulte, and Chrysanthos
  Dellarocas.
\newblock Credit scoring with social network data.
\newblock \emph{Marketing Science}, 35\penalty0 (2):\penalty0 234--258, 2015.

\bibitem[{World Bank}(2011)]{worldBank2011}
{World Bank}.
\newblock General principles for credit reporting, 2011.
\newblock URL
  \url{http://documents.worldbank.org/curated/en/662161468147557554/pdf/70193-2014-CR-General-Principles-Web-Ready.pdf}.

\bibitem[{World Bank}(2017)]{worldBank}
{World Bank}.
\newblock Financial inclusion, 2017.
\newblock URL
  \url{http://www.worldbank.org/en/topic/financialinclusion/overview#1}.

\end{thebibliography}

\end{document}